\documentclass[prd,preprintnumbers,twocolumn,eqsecnum,floatfix,letterpaper,superscriptaddress,nofootinbib]{revtex4}
\usepackage{color}
\usepackage[usenames,dvipsnames]{xcolor}
\usepackage{amsmath,amssymb,graphicx}
\usepackage{aas_macros}
\usepackage{bm}
\usepackage{times}
\usepackage{microtype}
\usepackage{booktabs}
\usepackage{subfigure}
\usepackage[normalem]{ulem}
\usepackage[varg]{txfonts}
\usepackage[colorlinks, pdfborder={0 0 0}]{hyperref}

\definecolor{LinkColor}{rgb}{0.75, 0, 0}
\definecolor{CiteColor}{rgb}{0, 0.5, 0.5}
\definecolor{UrlColor}{rgb}{0, 0, 0.75}
\hypersetup{linkcolor=LinkColor}
\hypersetup{citecolor=CiteColor}
\hypersetup{urlcolor=UrlColor}
\allowdisplaybreaks
\newcommand{\ICTS}{International Centre for Theoretical Sciences, Tata Institute of Fundamental Research, Bangalore 560012, India}
\newcommand{\Birmingham}{School of Physics and Astronomy, University of Birmingham, Edgbaston, Birmingham B15 2TT, United Kingdom}
\newcommand{\np}{\vec{\theta}}
\newcommand{\di}{\mathrm{d}}
\newcommand{\info}{\mathrm{I}}
\newcommand{\lal}{\textsc{LAL}}
\newcommand{\lalinf}{\textsc{LALInference}}
\newcommand{\lalinferencenest}{\textsc{LALInferencenest}}
\newcommand{\imrphenb}{\textsc{IMRPhenomB}}

\newcommand{\red}[1]{\textcolor{black}{{#1}}}
\def\emm{\color{magenta}}

\begin{document}

\title{Estimating parameters of binary black holes\\from gravitational-wave observations of their inspiral, merger and ringdown}
\author{Archisman Ghosh}
\affiliation{\ICTS}
\author{Walter Del Pozzo}
\affiliation{\Birmingham}
\author{Parameswaran Ajith}
\affiliation{\ICTS}
\begin{abstract}
We characterize the expected statistical errors with which the parameters of black-hole binaries can be measured from gravitational-wave (GW) observations of their inspiral, merger and ringdown by a network of second-generation ground-based GW observatories. We simulate a population of black-hole binaries with uniform distribution of component masses in the interval \red{$(3,80)~M_\odot$}, distributed uniformly in comoving volume, with isotropic orientations. From signals producing signal-to-noise ratio $\geq 5$ in at least two detectors, we estimate the posterior distributions of the binary parameters using the Bayesian parameter estimation code \lalinf. The GW signals will be redshifted due to the cosmological expansion and we measure only the ``redshifted'' masses. By assuming a cosmology, it is possible to estimate the gravitational masses by inferring the redshift from the measured posterior of the luminosity distance. We find that the measurement of the gravitational masses will be in general dominated by the error in measuring the luminosity distance. In spite of this, the component masses of more than 50\% of the population can be measured with accuracy better than \red{$\sim$25\%} using the Advanced LIGO-Virgo network. Additionally, the mass of the final black hole can be measured with median accuracy \red{$\sim 18\%$}. Spin of the final black hole can be measured with median accuracy \red{$\sim 5\% ~(17\%)$} for binaries with non-spinning (aligned-spin) black holes. Additional detectors in Japan and India significantly improve the accuracy of sky localization, and moderately improve the estimation of luminosity distance, and hence, that of all mass parameters. We discuss the implication of these results on the observational evidence of intermediate-mass black holes and the estimation of cosmological parameters using GW observations. 
\end{abstract}
\preprint{LIGO-P1500061-v2}
\maketitle

\section{Introduction and summary}

While we are nearing the centenary year of the first black hole solution in General Relativity (GR) discovered by Karl Schwarzschild~\cite{1916AbhKP1916..189S}, a number of astronomical observations have provided compelling, albeit indirect, evidence of the existence of astrophysical black holes (see~\cite{Narayan:2013gca} for a review). These observations strongly point to the existence of at least two populations of black holes: stellar-mass black holes with masses $\sim 5-30~M_\odot$ found in x-ray binaries and supermassive black holes with masses $\sim 10^6-10^{10}~M_\odot$ found in galactic nuclei~\cite{Narayan:2013gca}. There is also suggestive, but much less robust, evidence of a third population of black holes with intermediate ($\sim 10^2-10^3~M_\odot$) masses~(see,~e.g.,~\cite{Miller:2003sc,Pasham:2015tca}). 

The astronomical observations used to infer the existence of black holes are indirect. The mass measurements are based on observing the motion of luminous objects (such as stars or accreting matter) around a compact object. Furthermore, they only point to the existence of compact objects which are significantly more massive than other known compact objects, such as neutron stars. These observations are however inadequate to establish that these objects are indeed black holes predicted by General Relativity as opposed to more exotic compact objects, such as Boson stars. Additionally, these (indeed remarkable) measurements can be affected by systematic errors that are hard to track down, and hence are typically not included in the error estimates. Furthermore, the dynamical mass measurements from x-ray binaries only provide lower limits of the black hole's mass since what is measured is the \emph{mass function} of the black hole, which is a combination of the gravitational mass and inclination angle~\cite{Narayan:2013gca}.  

The anticipated gravitational-wave (GW) observations by the upcoming GW observatories provide a unique opportunity to directly measure the masses (and spin angular momenta) of black holes in coalescing black-hole binaries. Such binary systems are among the prime sources for the first direct detection of GWs using interferometric GW detectors, such as Advanced LIGO~\cite{0264-9381-27-8-084006}, Advanced Virgo~\cite{AdvVirgo:2009}, KAGRA~\cite{Somiya-Kagra:2012} and LIGO-India~\cite{LIGOIndiaProposal:2011}. The GWs, produced purely by the motion of the black holes in the binary and well described by the GR, travel to the detector completely unaffected by the intervening matter. By comparing the observational data with theoretical templates of the expected signals (as computed by GR) it is possible to extract the parameters of the binary (such as the masses and spin angular momenta of the black holes, sky-location and luminosity distance to the binary, etc.). Such observations, where the systematic errors are small compared to electromagnetic observations, are expected to provide the first direct measurements of the mass and spin angular momenta of black holes in the next decade. In principle, such observations will also provide us an opportunity for testing whether astrophysical black holes behave according to the black hole solutions predicted by GR~(see,~e.g.,~\cite{Yunes:2013dva}).

This paper aims to characterize the limiting statistical errors in the estimation of the parameters of coalescing binary black holes by advanced GW detectors. The recent advances in numerical relativity~\cite{Pretorius:2005gq,Baker:2005vv,Campanelli:2005dd} and analytical relativity~\cite{Blanchet:2013haa} have provided us with waveform templates that model the complete inspiral, merger and ringdown of the coalescence~\cite{Buonanno:2007pf,Pan:2011gk,Ajith:2007kx,Ajith:2007xh,Ajith:2009bn,Santamaria:2010yb,Taracchini:2012ig,Taracchini:2013rva,Damour:2014sva,Damour:2012ky,Damour:2007yf,Damour:2008te}. Previous estimates employing Fisher matrix formalism~\cite{Ajith:2009fz} have shown that this will significantly improve the parameter estimation accuracies of ``high-mass'' binaries as compared to estimates using templates only modeling the inspiral stage of the coalescence. Some of the more recent studies employing Bayesian parameter estimation codes have further demonstrated this~\cite{Littenberg:2012uj,Veitch:2015ela,Graff:2015bba}. 

Here we present a comprehensive study of the expected statistical errors in the context of the upcoming advanced GW detector networks. We simulate a population of black-hole binaries with uniform distribution of component masses in the interval \red{$(3,80)~M_\odot$}, distributed uniformly in comoving volume with isotropic orientations. The choice of the mass range is motivated by some of the recent population synthesis models~\cite{2012ApJ...759...52D}. From signals producing signal-to-noise ratio $\geq 5$ in at least two detectors, we estimate the posterior distributions of the binary parameters using the Bayesian parameter estimation code \lalinf~\cite{LALInference}. In addition to presenting error estimates in the parameters of the binary such as the component masses, we also investigate how well the mass and the spin of the newly formed black hole (as the product of the merger) can be estimated.

Figures~\ref{fig:err_vs_mtot} and \ref{fig:error_dists}, and Table~\ref{tab:summary_table} summarize the statistical errors with which various parameters can be measured from the simulated population of binary black holes. A brief summary of the main results is as follows: We find that the error in the measurement of the mass parameters are, in general, dominated by the error in the measurement of the luminosity distance to the source. This is due to the fact that the masses appear in the GW signal completely degenerate with the cosmological redshift $z$, and we can measure only the ``redshifted'' total mass  $M^z \equiv M (1+z)$. In order to estimate the real gravitational mass $M$, the redshift has to be estimated independently. In the absence of any independent measurement, $z$ can be computed from the measured luminosity distance if we assume a cosmology. Since luminosity distance is not a very well measured quantity, the errors in the measurements of the physical masses are dominated by the error in measuring the luminosity distance. In spite of this, the component masses of more than 50\% of the population of non-spinning can be measured with accuracy better than \red{$\sim$25\%} using the Advanced LIGO-Virgo three-detector network. Additionally, the mass and spin of the final black hole can be measured with median accuracy \red{$\sim 18\%$} and \red{$\sim 5\%$}. This will be of great interest to astrophysics since we are measuring the mass and spin of individual black holes (as opposed to combinations like chirp mass or total mass). If the black holes have non-precessing spins, the accuracies are slightly worse due to the correlation between mass and spin (in particular, the estimate of the final spin, which is reduced to $\sim 17\%$). 

As far are the redshifted parameters are concerned, as found in previous studies~\cite{Ajith:2009fz,Graff:2015bba,Veitch:2015ela}, we find that the total mass of the binary $M^z \equiv m_1^z+m_2^z$ is a better estimated mass parameter for heavier binaries (\red{$M^z \gtrsim 150 M_\odot$}), while the chirp mass $\mathcal{M}^z \equiv M^z \eta^{3/5}$ is the best estimated parameter for lighter binaries (\red{$M^z \lesssim 150 M_\odot$}), where $\eta \equiv m_1 m_2/M^2$ is the symmetric mass ratio of the binary. In addition, the mass $M_f^z$ of the final black hole is measured with remarkable accuracy. In fact, we find that for masses $M^z\gtrsim200M_\odot$, the mass of the final black hole $M^z_f$ is measured to a better precision than even the total mass $M^z$. In contrast, the mass of the individual black holes is measured rather poorly, limited by the measurement of the mass ratio.

Additional detectors in Japan and India significantly improve the accuracy of sky localization, and moderately improve the estimation of luminosity distance, and hence that of all mass parameters. The accuracy of estimating the physical mass parameters can be significantly improved by an independent measurement of $z$, for example from spectroscopic studies of the host galaxy, if the host galaxy can be identified. Alternatively, an improved measurement of the luminosity distance from GW observations{ (e.g., by reducing the correlation between the luminosity distance and the inclination angle by employing waveform templates including the effect of non-quadrupole modes, spin precession, etc.) can improve the accuracy of the mass parameters.  

The rest of the paper is organized as follows: Sec.~\ref{sec:methodology} presents details of the methodology used in this study, including the Bayesian parameter estimation pipeline, gravitational waveform models, the astrophysical setup, detector models, etc. Section~\ref{sec:results} presents a discussion of the results while Sec.~\ref{sec:end} discusses the implication of our results along with some concluding remarks. Throughout the paper we assume a flat $\Lambda$CDM cosmology with parameters $H_0=70\,\mathrm{km/s/Mpc}$, $\Omega_m=0.3$, $\Omega_\Lambda=0.7$, and geometric units: $G = c = 1$. 

\begin{figure*}
\centering
\includegraphics[width=0.99\textwidth]{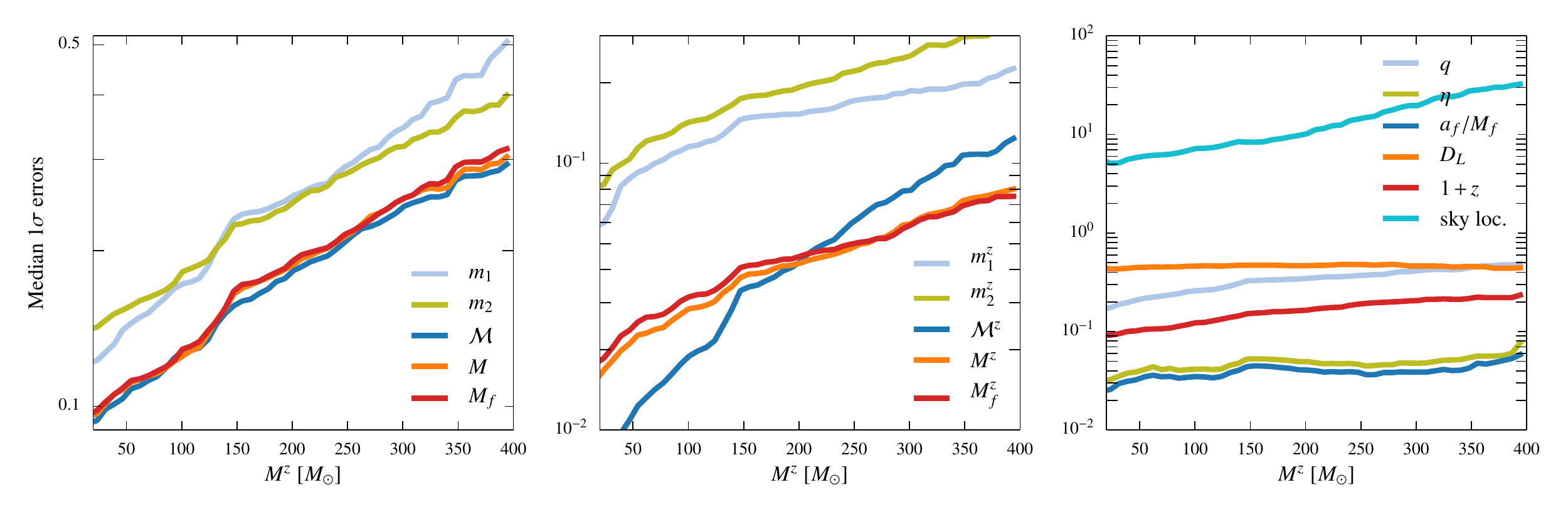}
\caption{Median values of the $1\sigma$ errors in estimating various parameters of non-spinning binary black holes using the three-detector (Advanced LIGO-Virgo) network as a function of the (redshifted) total mass of the binary. The error in sky location is in square degrees, while the rest are fractional errors. The error in estimating the physical mass parameters (left plot) are dominated by the error in estimating the luminosity distance. Hence they are estimated with poorer accuracy as compared to their redshifted counterparts. For the case of the redshifted mass parameters (middle plot), the chirp mass $\mathcal{M}^z$ is the best estimated mass parameter at low masses ($M^z\lesssim150M_\odot$), while at high masses ($M^z\gtrsim200M_\odot$) the total mass $M^z$ and the final mass $M_f^z$ are better estimated.}
\label{fig:err_vs_mtot}
\end{figure*}

\begin{table}
\centering
\begin{tabular}{c@{\quad}c @{\quad}c}
\toprule
Parameter & 3 detector & 5 detector \\
\midrule
Component masses $m_{1,2}$ & 24.5\% (41.9\%) & 22.6\% (40.0\%) \\
Chirp mass $\mathcal{M}$ & 17.4\% (29.9\%) & 15.4\% (26.6\%) \\
Total mass $M$ & 18.2\% (30.5\%) & 15.9\% (27.5\%) \\
Final mass $M_f$ & 18.3\% (31.0\%) & 16.1\% (27.8\%) \\
Mass ratio $q$ & 34.8\% (60.7\%) & 33.1\% (60.8\%) \\
Symmetric mass ratio $\eta$ & 5.4\% (16.0\%) & 5.5\% (16.1\%) \\
Final spin $a_f/M_f$ & 4.5\% (13.7\%) & 4.6\% (13.5\%) \\
Luminosity distance $d_L$ & 46.6\% (75.8\%) & 40.9\% (65.0\%) \\
Sky location $\Omega$ (sq. deg.) & 31.25 (121.34) & 8.90 (36.41) \\
\bottomrule
\end{tabular}
\caption{Summary of the parameter estimation capabilities of the upcoming GW detector networks. The table shows the 50 percentile (90 percentile) values of $1\sigma$ errors in estimating the parameters of the simulated binary black-hole population using the 3 detector network and the 5 detector network considered in the paper. Errors in all the parameters except the sky location are fractional errors. Here the initial black holes are assumed to have negligible spins.}
\label{tab:summary_table}
\end{table}


\section{Methodology}
\label{sec:methodology}

\subsection{Bayesian approach to parameter estimation}

In the presence of a GW signal, the strain of a GW detector $d(t)$ is given by 
\begin{align}
d(t) = n(t) + h(t)
\end{align} 
where $n(t)$ is the detector noise time series and $h(t)$ is the GW signal. 
As customary, we are assuming the detector noise to be well modelled
by a stationary Gaussian process with zero mean. Stationarity implies
that in the frequency domain all frequency components are independent
of each other, thus allowing us to write the autocorrelation of the
noise as
\begin{align}
\left< n(f) \, n^*(f^\prime) \right>=\frac{1}{2}S(f) \, \delta(f-f^\prime)
\end{align}
where we introduced the (one-sided) power spectral density $S(f)$. 
Given the above assumptions, the likelihood for a given noise
realization is given by 
\begin{align}
p(n|\info)\propto \exp\left[-\frac{(n|n)}{2}\right]
\end{align}
where we introduced the scalar product
\begin{align}
 (a|b)\equiv 2 \, \mathrm{Re}\int_0^\infty \di f\, \frac{a(f)b^*(f)+a^*(f)b(f)}{S(f)}\,.
\end{align}
The noise statistical properties allow us to write the likelihood of a
given detector strain realization given a GW signal as 
\begin{align}\label{eq:like}
p(d|h,\info)\propto \exp\left[-\frac{(d-h|d-h)}{2}\right]\,.
\end{align}
Assume that the GW signal $h(t)$ depends on some
parameters $\np$ which are unknown and we intend to infer from the
data. We can do so using Bayes' theorem: 
\begin{align}\label{eq:posteriors}
p(\np|d,h,\info)=p(\np|h,\info)\frac{p(d|\np,h,\info)}{p(d|h,\info)}
\end{align}
where $p(\np|h,\info)$ is the \emph{prior} probability distribution
for the parameters $\np$ given a GW model $h$,
$p(d|\np,h,\info)$ is the likelihood we introduced in
Eq.~(\ref{eq:like}) and $p(d|h,\info)$ is sometimes referred to as the
\emph{evidence} or marginal likelihood and it is given by
\begin{align}\label{eq:evidence}
p(d|h,\info) = \int \di\np\, p(\np|h,\info)p(d|\np,h,\info)\,.
\end{align}
Being able to infer the values of the parameters amounts to being able
to construct the posterior distribution $p(\np|d,h,\info)$. This is a
formidable task which can only be tackled numerically through stochastic
samplers. 
This is mainly due to the large dimensionality of the parameter space:
For non-spinning binary systems the vector is 9-dimensional, for more
complicated (and realistic models) the dimensionality rises quickly to
15 for fully spinning stellar systems to even larger integer once more
detailed physics is included in the gravitational waveform. 

For our purposes, we relied on the \lalinf\ \cite{LALInference} stochastic samplers available within the LIGO Algorithm Library (\lal) \cite{urlLAL}. In particular we made use of the \lalinferencenest\ software which implements a nested sampling algorithm \cite{Skilling2004a} in the context of GW data analysis. Nested sampling algorithms aim at solving the integral Eq.~(\ref{eq:evidence}) rather than producing samples from Eq.~(\ref{eq:posteriors}) and obtain posterior samples as a byproduct. We obtain posterior distributions on the following parameters this way: the total mass $M \equiv m_1+m_2$, mass ratio $q \equiv m_1/m_2$, time of arrival at geocenter $t_0$, phase of the waveform at a reference frequency $\varphi_0$, location of the binary on the sky $(\alpha, \delta)$, orientation of the binary w.r.t. the line of sight $(\iota, \psi)$, and the luminosity distance $D_L$. Most of the calculations in this paper are done assuming that the initial black holes have negligible spins. However, we do perform one set of simulation ascribing non-precessing spins to the initial black holes. For this case, the parameter space consists of two additional dimensions $\chi_1$ and $\chi_2$ corresponding to dimensionless spins of the two black holes (aligned/anti-aligned to the orbital angular momentum). Orbital eccentricity is assumed to be negligible.  Since the mass and the spin of the final black hole is uniquely predicted by the initial masses and spins, we can estimate the posteriors of the mass $M_f$ and dimensionless spin $a_f/M_f$ of the final black hole using fitting formulas calibrated to numerical-relativity simulations. For this work, we used the fitting formulas given in~\cite{Healy:2014yta}.

\subsection{Waveform models}

In this paper we characterize the expected statistical errors in the estimated parameters by employing the waveform family \imrphenb~\cite{Ajith:2009bn}. These waveforms describe the GW signals from the inspiral, merger and ringdown of binary black holes with non-precessing spins. The waveform is written in the frequency-domain as $h(f) \equiv A(f) \, e^{-\mathrm{i} \Psi(f)}$ where the amplitude $A(f)$ and the phase $\Psi(f)$  are defined as 
\begin{equation}
 A(f) \equiv \mathcal{C}\,\mathcal{A}
\begin{cases}
 f'^{-7/6} \left[ 1 + \sum_{i=2}^{3} \, \alpha_{i} \, v^{i} \right] &  f_0 \leq f < f_{1} \\
                                  w_{m} f'^{-2/3} \left[ 1 + \sum_{i=1}^{2} \, \epsilon_{i} \, v^{i} \right] & f_{1} \leq f < f_{2}\\
                                  w_{r} \mathcal{L} (f,f_{2},\sigma) & f_{2} \leq f < f_{3}, 
\end{cases}
\end{equation}
and 
\begin{equation}
 \Psi(f) \equiv 2\pi ft_{0} + \varphi_{0} + \frac{3}{128 \, \eta \, v^{5}}\left[ 1 + \sum_{k=2}^{7} \, \psi_k \, v^{k} \right]. 
\end{equation}
Above, 
\begin{eqnarray}
\mathcal{C} & \equiv & \frac{1}{2D_L} \, \sqrt{(1+\cos^2 \iota) \, F_+(\alpha, \delta, \psi) ^2 + 4 \cos^2 \iota \, F_\times(\alpha, \delta, \psi)^2}, \nonumber \\ 
\mathcal{A} & \equiv & \frac{{M}^{5/6} f_1^{-7/6}}{\pi^{2/3}} \sqrt{\frac{5\eta}{24}}
\end{eqnarray}
is the amplitude-scaling factor that depends on the antenna pattern functions $F_{+,\times}(\alpha, \delta, \psi)$ and the inclination angle $\iota$, where the angles $\alpha, \delta$ describe the sky-location of the binary and $\psi$ the polarization angle. Additionally, $f_0$ is the low-frequency cutoff of the detector noise where the PSD raises sharply due to seismic noise, $f_3$ is the high-frequency cutoff above which the power in the signal is negligible, $f_2$ and $f_3$ are the transition frequencies between the inspiral and the merger and between the merger and the ringdown. Also $f'\equiv f/f_1$ is a dimensionless frequency, $v \equiv (\pi M f)^{1/3}$, $\alpha_{2} = -323/224 + 451\eta /168$ and $\alpha_{3} = (27/8 - 11\eta/6)$ are the post-Newtonian (PN) corrections (1.5PN accurate) to the leading order amplitude of the inspiral, $\epsilon_{1} = 1.4547\chi - 1.8897$ and $\epsilon_{2} = -1.8153\chi + 1.6557$ are parameters describing the amplitude of the merger obtained from numerical-relativity simulations. $ \mathcal{L} (f,f_{2},\sigma)$ is a Lorentzian function with width $\sigma$ centered around $f_2$ while $w_{m}$ and $w_{r}$ makes the amplitude continuous over the transition frequencies $f_1$ and $f_2$. In the definition of the phase, $t_0$ is the arrival time of the signal at the detector, $\varphi_0$ is the phase at a reference frequency and $\psi_k$ are phenomenological parameters describing the phase evolution of the binary. The phenomenological parameters $\{f_1, f_2, f_3, \sigma, \psi_k\}$ are entirely functions of the physical parameters $\{M, \eta, \chi\}$ and are given in Table I of \cite{Ajith:2009bn}. Note that in this section, we did not include the effect of cosmological redshift in the observed waveform. This is described in the next section. 

\subsection{Astrophysical set up}
\label{sec:astro_setup}
\begin{figure}
\centering
\includegraphics[width=0.4\textwidth]{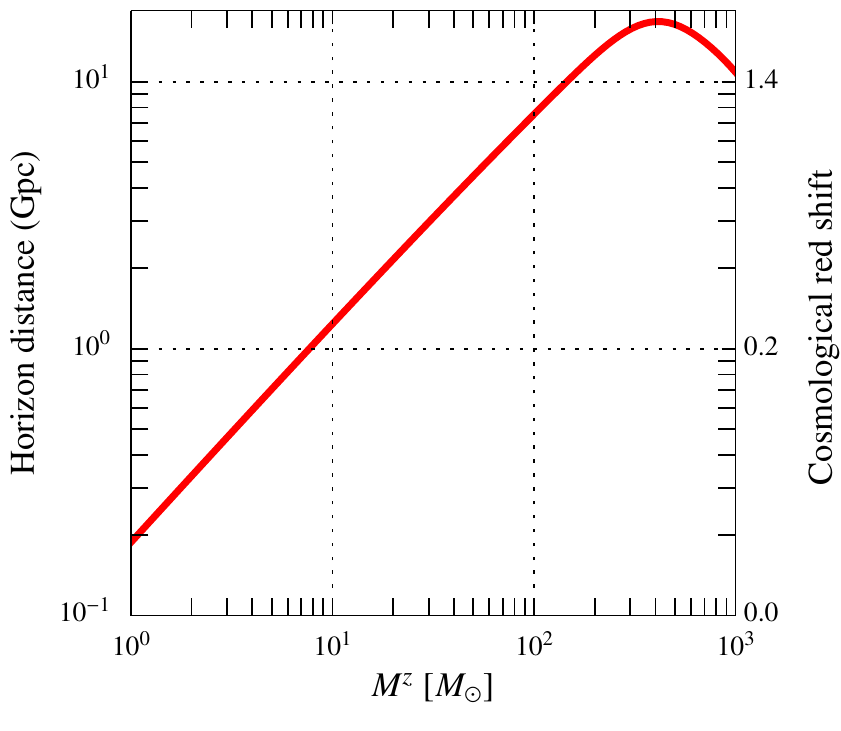}
\caption{The top plot shows the \emph{horizon distance} (distance to which optimally located- and oriented binaries can be observed with an optimal signal-to-noise ratio of 8) of the Advanced LIGO detector towards non-spinning, equal-mass binaries binaries as a function of the (redshifted) total mass of the binary. The bottom plot shows the expected distribution of SNRs from the population of binary black holes that cross the detection criterion.} 
\includegraphics[width=0.38\textwidth]{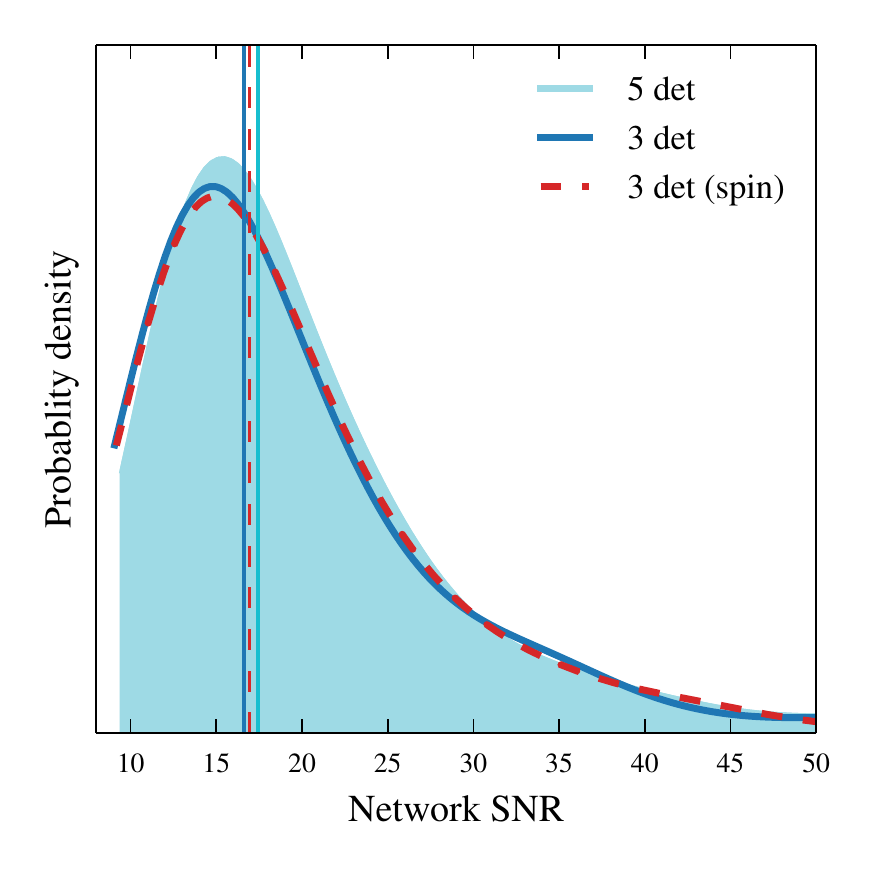}
\label{fig:horizon_and_snr}
\end{figure}

\begin{figure*}
\centering
\includegraphics[width=\textwidth]{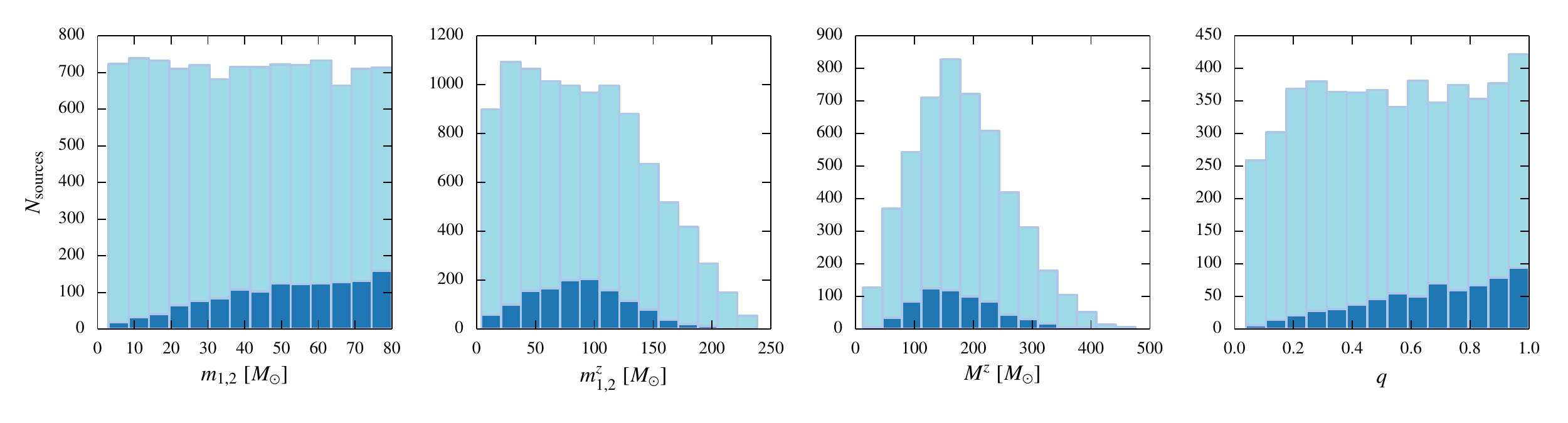}
\caption{The left plot shows the distribution of the rest-frame component masses $m_{1,2}$ of the simulated population of non-spinning binaries (light colored histograms) and that of the sub-population that crossed the detection threshold for the three-detector network (dark colored histograms). The two middle plots show the distributions of the redshifted component masses $m_{1,2}^z$ and the total mass $M^z$. Note that, even stellar-mass ($m_{1,2} < 80 M_\odot$) black holes can appear to have intermediate masses ($m^z_{1,2} > 100 M_\odot$) due to the degeneracy between mass and cosmological redshift. The right plot shows the distributions of the mass ratio $q \equiv m_2/m_1$. Note that comparable-mass $(q \gg 0)$ binaries are detected preferentially due their larger intrinsic luminosity.}
\label{fig:sim_and_det_distribution}
\end{figure*}

Our purpose is to understand the accuracy with which we can realistically expect to measure the parameters of binary black holes from GW
observations of high-mass stellar black holes. For this purpose we simulate a population of \red{$5000$} binaries with rest-frame component masses uniformly distributed in \red{$(3,80)~M_\odot$}. The binaries are also distributed uniformly in co-moving volume, therefore uniform on the celestial 2-sphere and in redshift according to the probability distribution \cite{Coward2005a}
\begin{align} \label{eq:pz1}
	p(z) = \frac{\di R(z)}{\di z}\frac{1}{R(z_{\mathrm{max}})}
\end{align}
in the redshift $z$ space. In Eq.~(\ref{eq:pz1}), $R(z)$ is the cosmic rate
whose evolution with redshift is equal to
\begin{align} \label{eq:pz2}
	\frac{\di R(z)}{\di z} = \frac{\di V}{\di z} \, \frac{r_0 \, e(z)}{1+z}
\end{align}
where $r_0$ is the rate of binary-black-hole coalescence in the local Universe, $e(z)$ is the cosmic star formation rate and $V$
is the co-moving volume. 
In a Friedmann-Robertson-Walker-Lema\^itre universe
\begin{align} \label{eq:pz3}
	\frac{\di V}{\di z} = 4\pi\frac{D_L^2(z)}{(1+z)^2 \, H(z)}
\end{align}
where 
\begin{align}
	H(z)=H_0\sqrt{\Omega_m (1+z)^3+\Omega_k (1+z)^2+\Omega_\Lambda}
\end{align}
is the Hubble parameter. Above, $H_0$ is the Hubble constant, $\Omega_m$ is the fractional matter density, $\Omega_\Lambda$ is the fractional energy density due to a cosmological constant, $\Omega_k=1-\Omega_m-\Omega_\Lambda$ is the fractional curvature energy density, which we take to be zero $(\Omega_k=0)$. The luminosity distance is then given by \cite{Hogg1999a}
\begin{align} \label{eq:dl}
	D_L(z)= (1+z)\int_0^z\frac{dz^\prime}{H(z^\prime)}. 
\end{align}
Since we are not interested in predictions relative to the coalescence rates of the high mass systems, we considered here the case $r_0 =1$ and $e(z) = \mathrm{const}$. 

It is worth noting that the mass parameters and the redshift are in general completely degenerate \cite{SathyaSchutz2009}, thus what is
observed in the detector is not the rest-frame mass, but the ``redshifted'' one. Therefore, the mass distribution of all detected sources will be quite different from the intrinsic distribution (see Fig.~\ref{fig:sim_and_det_distribution}). We will denote the redshifted mass parameters with a superscript $z$ ($m_{1,2}^z, M^z, \mathcal{M}^z, M_f^z$, etc.).

For the same set of masses, sky position, distances, orientation angles, we considered two scenarios, one where the spins of the BHs
are negligible and one where the dimensionless spins can assume take arbitrary values in the interval $(-1,1)$, but are restricted to lie in a direction aligned/anti-aligned to the orbital angular momentum, so that the spins do not precess. We did not consider the general case of fully precessing binary systems as no fully precessing waveform including also merger and ringdown is available yet and, considering the mass range we are interested in, merger and ringdown contribute significantly if not predominantly to the accumulation of signal-to-noise ratio in the detector. 

\begin{figure*}
\centering
\includegraphics[width=0.95\textwidth]{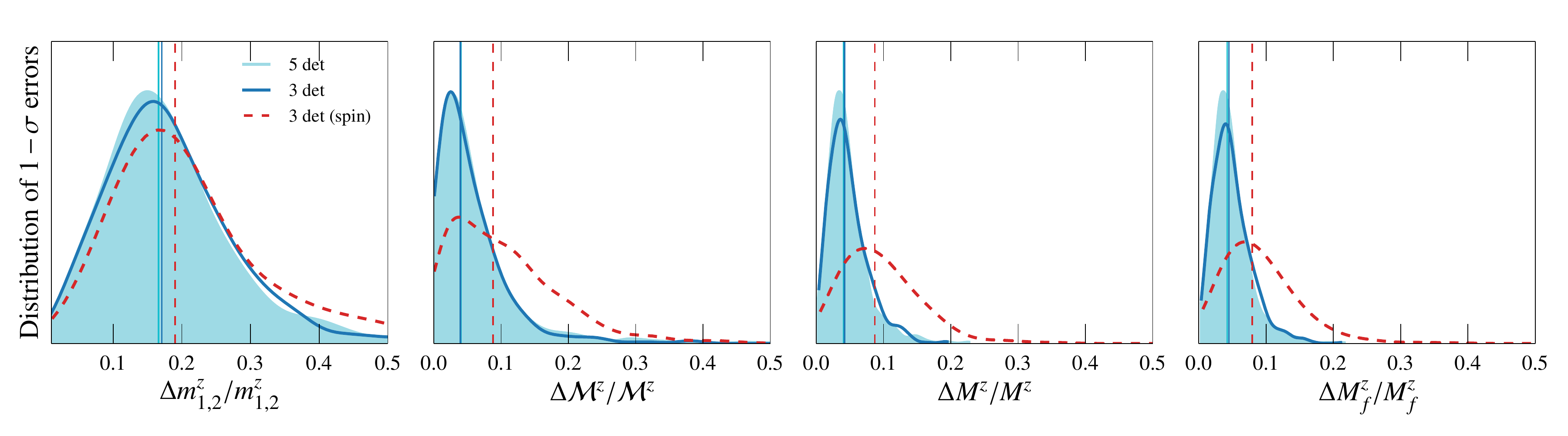}
\caption{The top panel shows the distribution of the $1\sigma$ errors in estimating the ``redshifted'' mass parameters (redshifted component masses $m^z_{1,2}$, chirp mass $\mathcal{M}^z$, total mass $M^z$ and the final mass $M^z_f$) from the simulated population of coalescing binary black holes. In each subplot, the three distributions correspond to non-spinning binary black holes detected by the 3-detector network and the 5-detector network that we consider, and the population of binary black holes with non-precessing spins detected by the 3-detector network. The vertical lines show the median of the distribution. The bottom panels show the distribution of the errors in estimating the ``physical'' parameters.  It can be seen that the error in measuring the ``physical'' mass parameters $\mathcal{M}$, $M$ and $M_f$ are significantly larger than that in measuring the ``redshifted'' parameters, and are dominated by the error in measuring the luminosity distance. The error in the component masses are dominated by that in measuring the mass ratio of the system, and hence are not very different in the top and bottom panels. Note that the mass ratio $\eta$, the final spin $a_f/M_f$, the luminosity distance $d_L$ and the sky location $\Omega$ are independent of the cosmological redshift.} 
\includegraphics[width=0.95\textwidth]{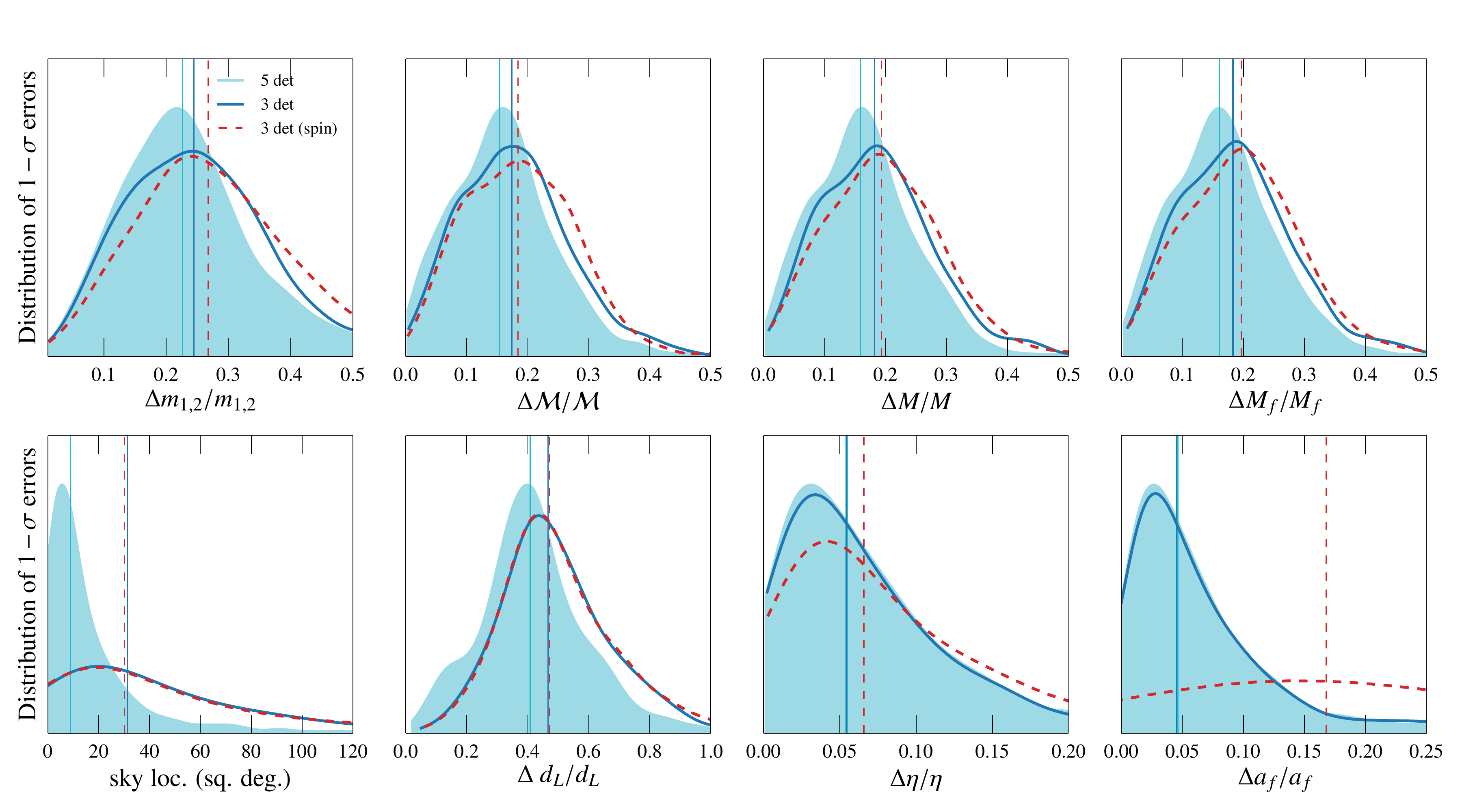}
\label{fig:error_dists}
\end{figure*}

\subsection{Detector models}

We consider two upcoming/anticipated networks of advanced GW observatories: 1) a three-detector network consisting of two Advanced LIGO detectors located in Hanford, WA and Livingston, LA in the United States and one Advanced Virgo network located in Cascina, Italy. 2) a five-detector network consisting of the three Advanced LIGO-Virgo detectors mentioned above and the KAGRA detector located in Kamioka, Japan and the LIGO-India detector located in India. Since the location of the LIGO-India is not finalized, we assume a fiducial location north of Bangalore~\cite{2011CQGra..28l5023S}. The Advanced Virgo detector is assumed to have the sensitivity given in~\cite{TheVirgo:2014hva} while the rest of the detectors are assumed to be having the sensitivity of the Advanced LIGO in the ``high-power-zero-detuning'' configuration~\cite{AdvLIGOPSD}. We assume a low-frequency cutoff of 10 Hz below which the detector has extremely poor sensitivity due to seismic noise. 

\subsection{Simulations}

From the simulated population of binary black holes described in Sec.~\ref{sec:astro_setup}, we select those binaries producing signal-to-noise ratio $\geq$ 5 in at least two detectors. This is to simulate a realistic detection scenario with two-detector coincidence. The distribution of the network signal-to-noise ratio of these ``detected'' binaries is shown in the bottom panel of Fig.~\ref{fig:horizon_and_snr}. The top panel of this figure shows the \emph{horizon distance} (the distance at which the detector can observe an optimally-located- and oriented binary with signal-to-noise ratio of 8) of Advanced LIGO towards non-spinning equal-mass binaries, which gives an idea of the distance reach of the detector network.  Posterior distributions of all the binary parameters are computed using the Bayesian parameter estimation code \lalinferencenest. We choose prior distributions as follows: observed component masses uniform in the interval $(2,300)~M_\odot$, sky location and orientation uniform on the 2-sphere, a uniform distribution for the time of arrival at geocenter $t_0$ with a width of 0.1 s around the true arrival time, phase $\phi_0$ uniformly distributed between $0$ and $2\pi$ and for the luminosity distance we assume a prior distribution given by Eqs.~(\ref{eq:pz1}), (\ref{eq:pz2}), (\ref{eq:pz3}) with $D_L \in (1,50000)$ Mpc. For the spinning case, we assume a uniform prior on the dimensionless spins for both components between $[-1,1]$. From the marginalized posterior of each parameter of interest, we compute the $1\sigma$ confidence intervals. 

\section{Results and discussion}
\label{sec:results}

\begin{figure*}
\centering
\subfigure{\includegraphics[width=0.40\textwidth]{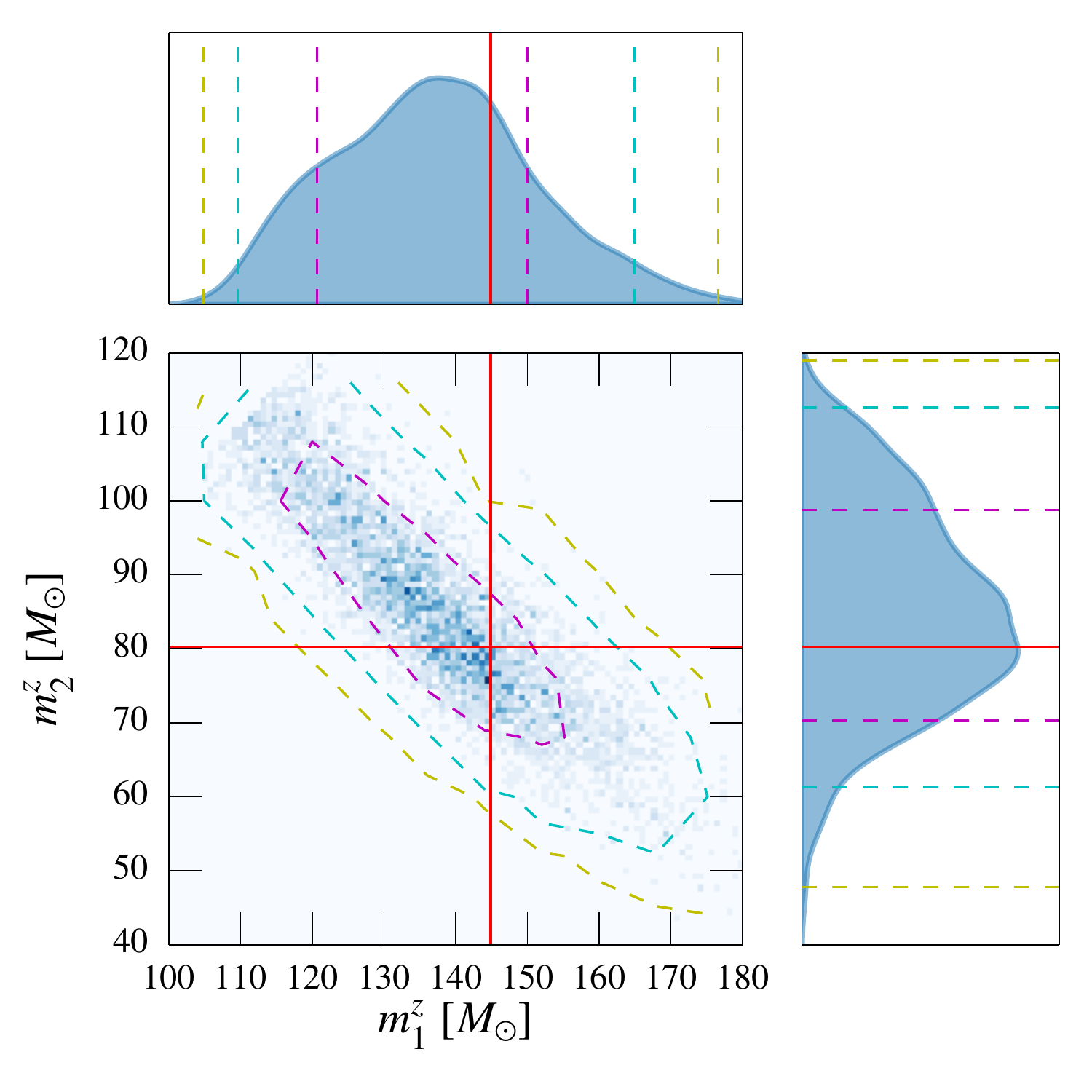}}
\subfigure{\includegraphics[width=0.40\textwidth]{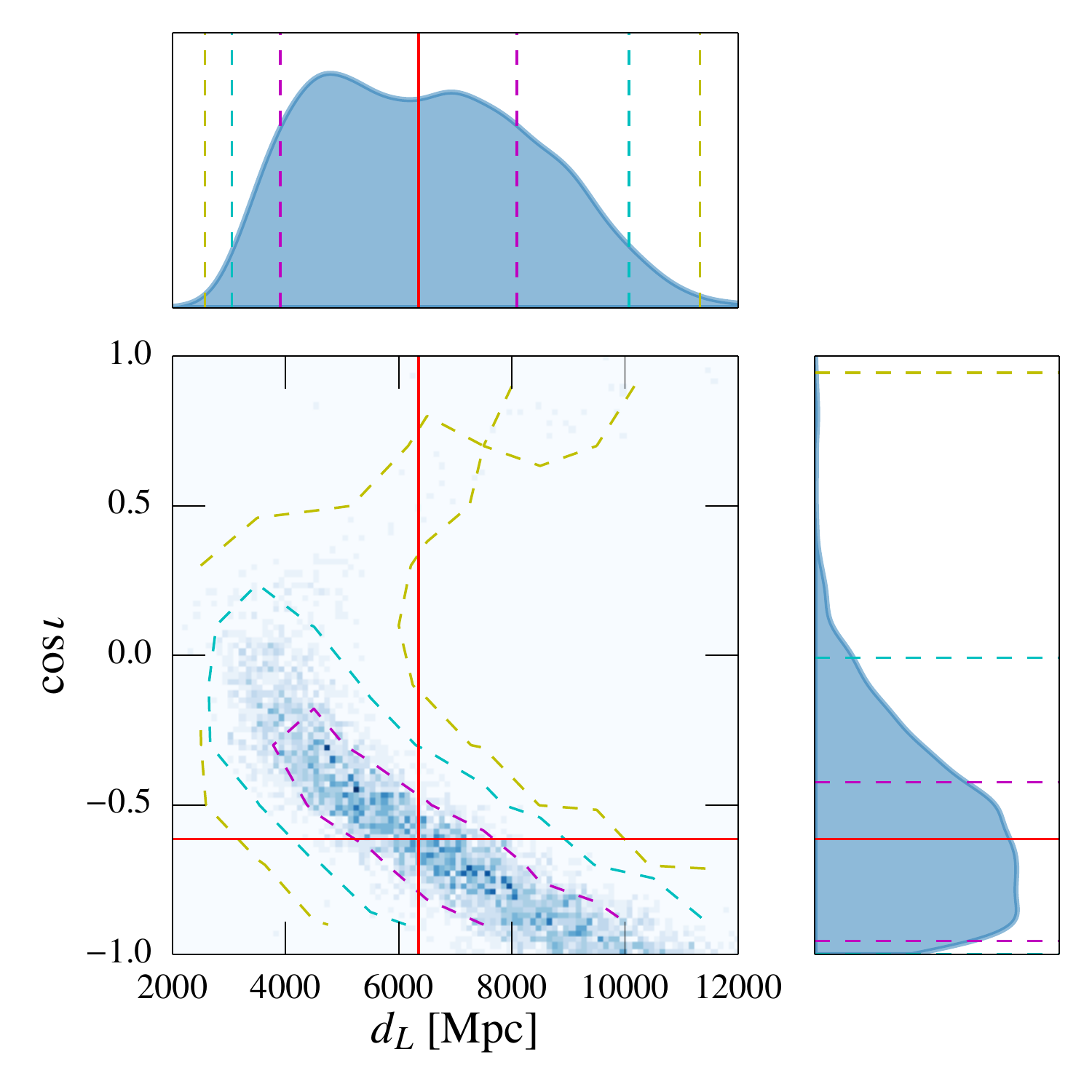}}
\caption{Typical posteriors obtained using \lalinf. The solid red lines indicate the injection values. The dashed magenta, cyan and yellow contours indicate the 68.3\%, 95.5\% and 99.7\% confidence regions respectively. The left panel shows the correlation between the component masses. For large masses, the total mass is roughly the most precisely estimated quantity. The posterior samples are thus correlated along the $M^z=\rm{constant}$ line. The right panel shows the correlation between luminosity distance and inclination. In this case $d_L/(1+\cos^2\iota)$ is the precisely estimated quantity. Because of this degeneracy, neither of the two physical quantities are well-estimated.}
\label{fig:corr_pos}
\end{figure*}

The main results of our work are summarized in Figs.~\ref{fig:err_vs_mtot} and ~\ref{fig:error_dists}, and in Table~\ref{tab:summary_table}. To remind the reader, we simulate an astrophysical simulation of \red{5000} black holes with component masses uniformly distributed between \red{$(3,80)~M_\odot$}. Figure~\ref{fig:err_vs_mtot} shows the median value of the $1\sigma$ errors on various estimated parameters as a function of the (redshifted) total mass $M^z$ of the binary. Consistent with our expectations, we find that the chirp mass $\mathcal{M}^z$ is the most precisely estimated quantity for low masses $M^z\lesssim150M_\odot$, the total mass $M^z$ best estimated for intermediate masses $150M_\odot\lesssim{M^z}\lesssim200M_\odot$, while the mass of the final black hole $M_f^z$ is the best estimated mass parameter for high masses $M^z\gtrsim200M_\odot$. Figure~\ref{fig:error_dists} shows the distributions of the expected $1\sigma$ statistical errors on the estimated quantities from the simulated source population. The top panel shows the distributions of the ``redshifted'' parameters and the bottom panel shows the distribution of the ``physical'' parameters. The vertical lines of the same colors indicate the median of the respective distributions. A verbal summary of the results is as follows: 

\paragraph{Measurement of redshifted parameters:} For the case of non-spinning binaries, the chirp mass $\mathcal{M}^z$, the total mass $M^z$, and the final mass $M_f^z$ are estimated to better than 4\% (4\%) for more than half of the population, using a five (three) detector network. In contrast, the component masses $m_{1,2}$ are estimated to about 16\% (17\%). In the case of aligned-spin binaries, the measurement accuracies deteriorate, due to correlations between mass and spin. The median measurement accuracies for $m_{1,2}^z$, $\mathcal{M}^z$, $M^z$, $M_f^z$ are 19\%, 9\%, 9\%, 8\% using a three-detector network.

\paragraph{Measurement of ``physical'' parameters:} For the case of non-spinning binaries, the chirp mass $\mathcal{M}$, the total mass $M$, and the final mass $M_f$ are estimated to better than 16\% (18\%) for more than half of the population, using a five (three) detector network. Component masses $m_{1,2}$ are estimated to about 23\% (25\%). For aligned-spin binaries, the median measurement accuracies for $m_{1,2}$, $\mathcal{M}$, $M$, $M_f$ are 25\%, 18\%, 19\%, 20\% using a three-detector network. 

The sky location, as expected, shows a significant improvement with $5$ detectors as compared to $3$ detectors. This is because the leading estimate of the sky location comes from the difference in the time of arrival of the signal at the different detectors, and with $3$ detectors, in many cases, one cannot distinguish between two antipodal points in the sky. The estimate on the luminosity distance is quite imprecise as compared to the redshifted masses above --- the median error on distance is about $40\%$ and $50\%$ with $5$ and $3$ detectors respectively. Since the distance is used to calculate the redshift in order to translate from the observed to the physical masses, the errors in the physical chirp, total, and final masses are dominated by the inaccuracy of the distance estimate, and are much worse compared to their corresponding redshifted values. On the other hand, the mass ratio is unaffected by redshift. The errors in the component masses do not change much between their redshifted and physical values, since they are dominated by the error in estimating the mass ratio. Moreover, since the final spin $a_f/M_f$ is determined entirely by the mass ratio and the initial (dimensionless) spins, the final spin is also unaffected by redshift. 

We see that allowing for the possibility of aligned-spins, in general, deteriorates the estimation of mass and spin parameters. This is due to the correlation between the mass and the spins. This degeneracy is particularly pernicious for the case of templates with aligned spins~\cite{Baird:2012cu}, which could be improved by including the effect of spin precession~\cite{Chatziioannou2015}. The measurement of the physical mass masses, $M$,$\mathcal{M}$, and $M_f$ can be improved by improving the measurement of luminosity distance. Luminosity distance is not a well-estimated quantity, mainly because of its correlation with the inclination angle (see Figure~\ref{fig:corr_pos}). The degeneracy can be slightly broken and the estimates made more precise by including more detectors. We already see that going from $3$ to $5$ detectors improves the distance estimate by a factor of about {\emm $1.5$}. Including higher harmonics in the waveform can be even more effective in breaking this degeneracy, and can be of great importance in the measurement of the physical masses of black holes.

Let us also list some of the caveats of our results: Note that we used the analytical IMR gravitational-waveform family \imrphenb\ that has been calibrated against numerical-relativity simulations with mass ratio $q \geq 1/4$. However, the effect of any waveform systematics is likely to be very small: As seen in the right plot of Fig.~\ref{fig:sim_and_det_distribution}, the population of detected binaries is significantly dominated by those with comparable masses (due their larger intrinsic luminosity). Additionally, in our simulation we use a uniform distribution of component spins in the interval (-1,1), while \imrphenb\ has been calibrated against numerical-relativity waveforms with moderate spins ($|\chi_{1,2}| \leq 0.85$ for $q = 1$ and $|\chi_{1,2}| \leq 0.75$ for $q \leq 1$). This can introduce non-negligible systematics in the case of highly spinning binaries. Even though it is unlikely that this will significantly change the median values of the statistical errors that we quote, the error estimates for spinning binaries need to be looked at keeping this caveat in mind.

\section{Implications of the results}
\label{sec:end}

\paragraph*{Measuring black-hole mass- and spin functions:---} We find that the errors in measuring the ``physical'' mass parameters is in generally dominated by the error in measuring the luminosity distance, which is often overlooked. Even when taking this effect into account, we see that it is possible to measure the component masses of more than $\sim50\%$ of the assumed population with accuracies better than 25\%. We also find that the mass of the final black hole can be measured with a slightly better median accuracy of $\sim18\%$. More interestingly, the spin of the final black hole can be measured with a remarkable accuracy of $\sim 5-17\%$. This has to be combined with the fact that GW-based measurements will have significantly smaller systematic errors as compared to the electromagnetic measurements.

\paragraph*{Detecting intermediate-mass black holes:---} One of the interesting scientific potential of GW observations of binary black holes is to shed light on the possible existence of intermediate-mass black holes~\cite{Graff:2015bba,Veitch:2015ela}. Note that, due to cosmological redshift, even stellar-mass black holes ($m_{1,2} \lesssim 80 M_\odot$) can have observed mass greater than $100 M_\odot$ (see Fig.~\ref{fig:sim_and_det_distribution}). Hence the mass measurements from GW observations have to be interpreted carefully. However, it is possible to measure the physical component masses with reasonable accuracy ($\sim 25\%$). More interestingly, since we are able to measure the mass of the final black hole well, GW observations might enable us to witness the birth of intermediate-mass black holes through mergers. This would help shed light on the the yet unclear channels of formation.    

\paragraph*{Measuring cosmological parameters:---} It is, in principle, possible to estimate some of the cosmological parameters (such as $H_0$ and $\Omega_M$) by combining the estimates of the luminosity distance with independent estimates of the cosmological redshift from spectroscopic observations of the host galaxy~\cite{1986Natur.323..310S}. Using the median error estimates of the luminosity distance and sky location ($\sim 50\%$ and 30 sq. deg for the three-detector network), we estimate that \red{$\sim 200$} potential host galaxies are likely to be present in the allowed confidence region. This will be considerably reduced in the case of a five-detector network. Still, it will be impossible to uniquely identify the host galaxy. However, reasonable constraints on the cosmological parameters can be obtained by combining multiple GW observations~\cite{2012PhRvD..86d3011D}. This is being explored in detail in an ongoing work~\cite{Ghosh:2015cosm}. 

\acknowledgments
We thank Alberto Sesana, Alberto Vecchio and John Veitch for useful discussions. AG and PA acknowledge support from the AIRBUS Group Corporate Foundation through a chair in ``Mathematics of Complex Systems'' at the International Centre for Theoretical Sciences (ICTS). PA's research was supported by a Ramanujan Fellowship from the Science and Engineering Research Board (SERB), India, the SERB FastTrack fellowship SR/FTP/PS-191/2012, and by the Max Planck Society and the Department of Science and Technology, India through a Max Planck Partner Group at ICTS. Computations were performed using the ICTS computing clusters Mowgli and Dogmatix. The work was funded in part by a Leverhulme Trust research project grant. This paper has the LIGO document number LIGO-P1500061-v2.

\bibliography{ParamEstim}

\end{document}